\documentclass[aps,prl,reprint,amsmath,showpacs,superscriptaddress]{revtex4-1}
\usepackage{graphicx}
\usepackage{textcomp}
\usepackage{color}

\begin{document}

\title{Impact of thermal annealing on graphene devices encapsulated in hexagonal boron nitride}
	
\author{S.~Engels}
    \affiliation{JARA-FIT and 2nd Institute of Physics, RWTH Aachen University, 52074 Aachen, Germany}
    \affiliation{Peter Gr{\"u}nberg Institute (PGI-9), Forschungszentrum J{\"u}lich, 52425 J{\"u}lich, Germany}

\author{B. Terr\'es}
    \affiliation{JARA-FIT and 2nd Institute of Physics, RWTH Aachen University, 52074 Aachen, Germany}
    \affiliation{Peter Gr{\"u}nberg Institute (PGI-9), Forschungszentrum J{\"u}lich, 52425 J{\"u}lich, Germany}

\author{F. Klein}
    \affiliation{JARA-FIT and 2nd Institute of Physics, RWTH Aachen University, 52074 Aachen, Germany}

\author{S. Reichardt}
    \affiliation{JARA-FIT and 2nd Institute of Physics, RWTH Aachen University, 52074 Aachen, Germany}

\author{M.~Goldsche}
    \affiliation{JARA-FIT and 2nd Institute of Physics, RWTH Aachen University, 52074 Aachen, Germany}
    \affiliation{Peter Gr{\"u}nberg Institute (PGI-9), Forschungszentrum J{\"u}lich, 52425 J{\"u}lich, Germany}

\author{S. Kuhlen}
    \affiliation{JARA-FIT and 2nd Institute of Physics, RWTH Aachen University, 52074 Aachen, Germany}

\author{K.~Watanabe}
    \affiliation{National Institute for Materials Science, 1-1 Namiki, Tsukuba 305-0044, Japan}

\author{T.~Taniguchi}
    \affiliation{National Institute for Materials Science, 1-1 Namiki, Tsukuba 305-0044, Japan}

\author{C.~Stampfer}
    \affiliation{JARA-FIT and 2nd Institute of Physics, RWTH Aachen University, 52074 Aachen, Germany}
    \affiliation{Peter Gr{\"u}nberg Institute (PGI-9), Forschungszentrum J{\"u}lich, 52425 J{\"u}lich, Germany}


\begin{abstract}
We present a thermal annealing study on single-layer and bilayer (BLG) graphene encapsulated in hexagonal boron nitride. The samples are characterized by electron transport and Raman spectroscopy measurements before and after each annealing step. While extracted material properties such as charge carrier mobility, overall doping, and strain are not influenced by the	annealing, an initial annealing step lowers doping and strain variations and thus results in a more homogeneous sample. Additionally, the narrow 2D-sub-peak widths of the Raman spectrum of BLG, allow us to extract information about strain and doping values from the correlation of the 2D-peak and the G-peak positions.
\end{abstract}

\maketitle

\textit{Introduction} Within the past 10 years extensive research in the field of graphene has led to a number of discoveries of interesting physics such as the anomalous quantum Hall effect \cite{Zhang2005Nov,Novoselov2006Feb}, Klein tunneling \cite{Katsnelson2006Aug}, and Hofstadter butterfly physics in graphene-hexagonal boron nitride (hBN) heterostructures \cite{Dean2013May,Hunt2013Jun,Ponomarenko2013May}. Moreover, the increasing interest in graphene is not limited to fundamental research but extends to applications such as high frequency transistors \cite{Lin2010Feb}, flexible electronics \cite{Kim2009Jan}, or optoelectronics \cite{Bonaccorso2010Aug}. The rapid advancement in all those fields is strongly tied to the improvement of the sample quality since higher quality samples reveal the unperturbed exceptional properties of intrinsic graphene. A brief review of the development of the graphene sample quality since its first isolation in 2004 \cite{Novoselov2004Oct} is provided in Fig. 1. In particular, we show the charge carrier mobility (serving as a figure of merit for the quality) of single-layer (SLG), Fig. 1a and bilayer graphene (BLG), Fig. 1b within the past decade. In the plot we focus on the three most commonly used techniques to fabricate graphene devices for transport studies, i.e. graphene on SiO$_2$, suspended graphene, and graphene on hBN. In early days, SLG and BLG were mostly placed on SiO$_2$ substrates. In this case, the rough surface, charge traps, and the presence of dangling bonds \cite{Geringer2009Feb} limit the device quality. Consequently, charge carrier mobilities of only a few 10,000 cm$^2$/(Vs) have been reported. A straightforward method to overcome the bad influence of the substrate is to fabricate suspended graphene devices. This technique results in higher mobility samples for SLG and BLG with values up to $2 \times 10^6$~cm$^2$/(Vs) \cite{Mayorov2012Sep}. However, as a considerable disadvantage, the suspended graphene devices suffer from the fact that these structures cannot be fabricated in arbitrary geometries and are extremely fragile and sensitive to the environment. All these limitations can be overcome by supporting and encapsulating graphene with atomically flat hBN \cite{Dean2010Aug,Wang2013Nov}. In this way, graphene can be protected from any environmental disturbances, including ad-atoms. Moreover, the flat surface of the hBN limits buckling induced effects. Consequently, hBN has proven to be a valuable substrate allowing the carrier mobility in graphene to reach values of up to 1$\times$10$^6$cm$^2$/(Vs) \cite{Wang2013Nov}. In this work, we show experimental evidence that encapsulated graphene can withstand substantial heat treatment without any decrease in sample quality. We show that thermal annealing does only slightly modify the electronic properties of these structures. In particular, we investigate the carrier mobility and the Raman spectra of SLG-hBN heterostructures while annealing the sample with maximum temperatures up to $T_{max}$ = 525$^\circ$C. We observe that a very first annealing step helps to make the sample more homogenous while all investigated properties stay unchanged after further annealing. Additionally, the study has been extended to BLG, where we show (i) that due to the narrow width of the 2D-sub-peaks, strain values can be extracted by the so-called vector decomposition method \cite{Lee2012Aug}, and (ii) that annealing leads to a slightly increased homogeneity in the sandwiched BLG in agreement with the SLG study. Our findings make these graphene heterostructures particularly interesting for electronic applications in an environment exhibiting extreme thermal conditions.

\begin{figure*}	
\includegraphics[width=0.95\linewidth]{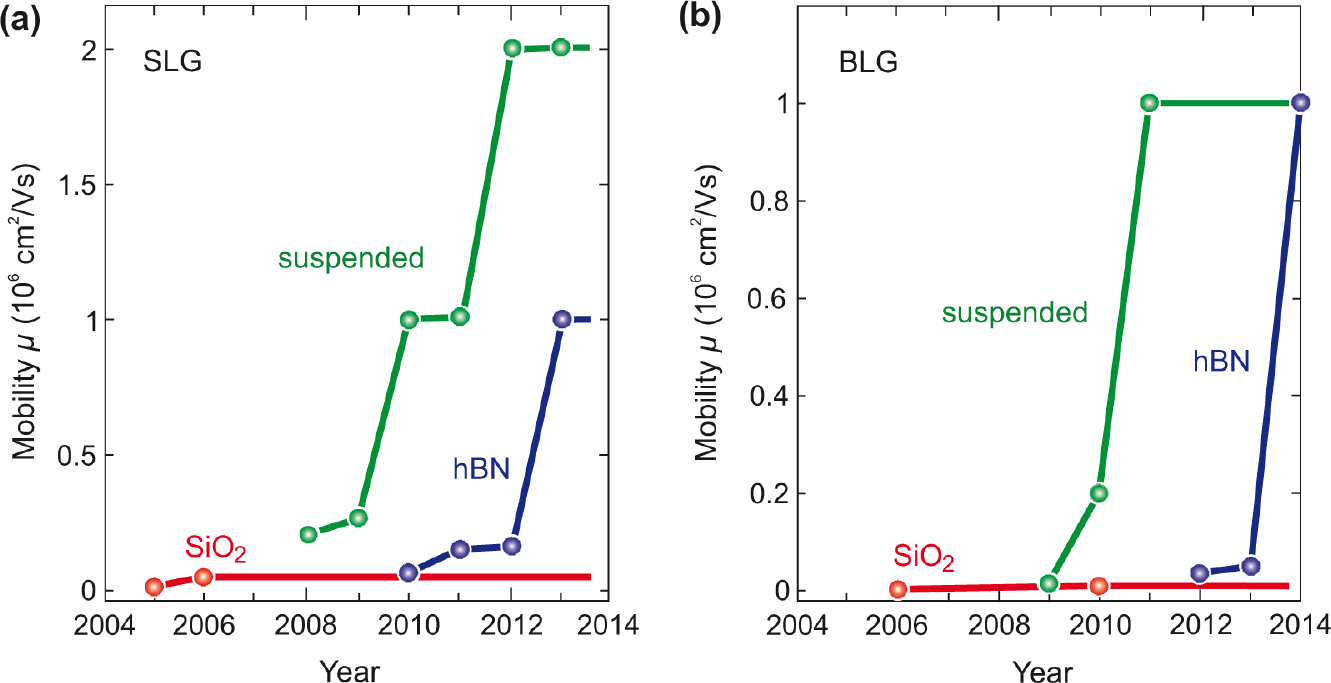}
\caption{
        Maximum mobility of single-layer (SLG) (a) and bilayer (BLG) (b) graphene from 2004 to 2014 for devices based on SLG/BLG on SiO$_2$, on hBN and suspended SLG/BLG devices. The data points are extracted from literature, where for SLG Refs. \cite{Zhang2005Nov,Zhang2006Apr} served as references on SiO$_2$, Refs. \cite{Bolotin2008Jun,Du2009Oct,Castro2010Dec,Mayorov2012Sep} as references for suspended structures and Refs. \cite{Dean2010Aug,Abanin2011Apr,Wang2013Nov} for hBN. In case of BLG, we extracted the mobility values from Refs. \cite{Novoselov2006Feb,Zhao2010Feb} for SiO$_2$, Refs. \cite{Feldman2009Sep,Bao2010Dec,Mayorov2011Aug} for suspended structures, and Refs. \cite{Goossens2012Sep,Maher2014Jul} for hBN.
}
\label{fig:MPRs}
\end{figure*}

\begin{figure}	
\includegraphics[width=\columnwidth]{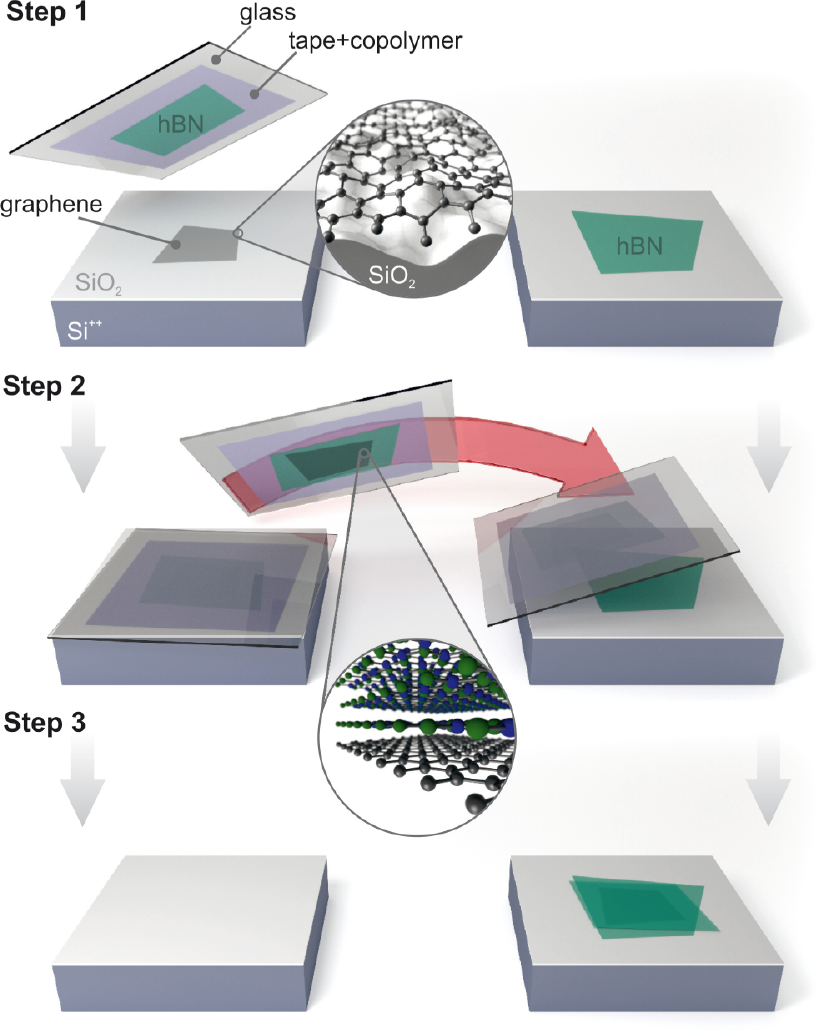}
\caption{
        Illustration of the fabrication process of hBN-graphene-hBN heterostructures, which is based on the process discussed in Ref. [19]. In \textit{step 1} graphene and hBN flakes are prepared on highly doped silicon substrates covered with a thermal oxide (Si$^{++}$/SiO$_2$). Additionally, an hBN flake is prepared on a glass/adhesive tape/copolymer stack. In \textit{step 2} the graphene is transferred onto the hBN on SiO$_2$ by the glass stack. Finally, in \textit{step 3} the glass stack is removed and the hBN-graphene-hBN heterostructure is cleaned by organic solvents.
}
\label{fig:MPRs}
\end{figure}

\begin{figure}	
\includegraphics[width=\columnwidth]{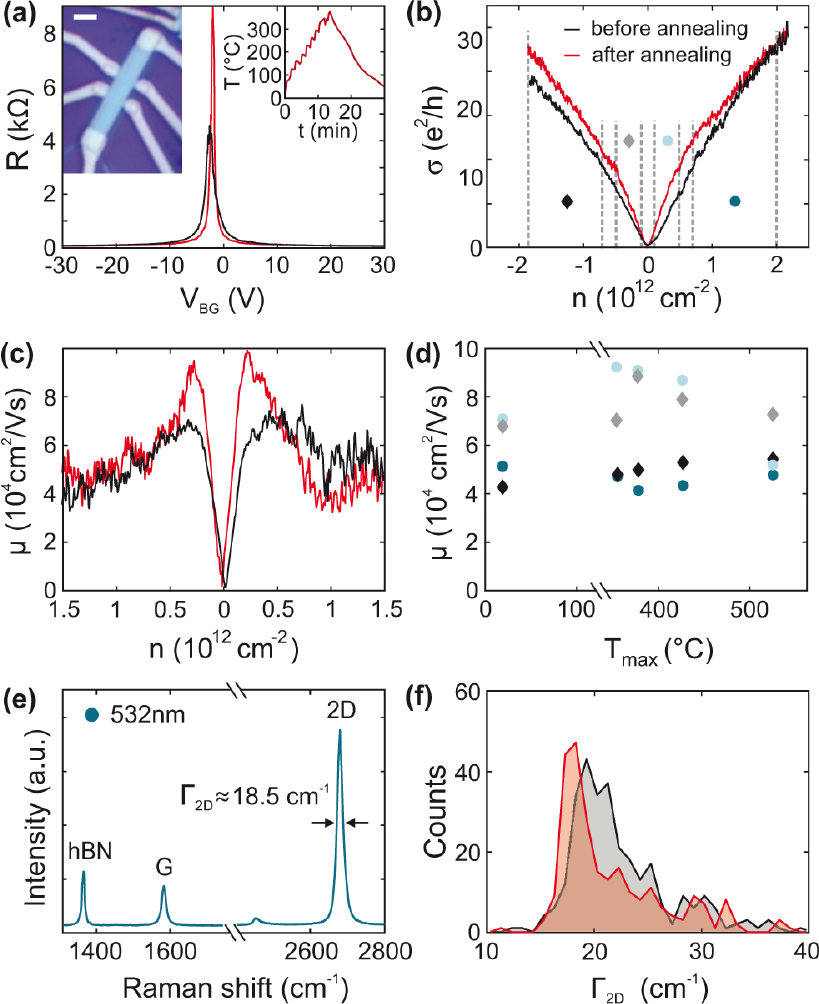}
\caption{
        (a) Four-terminal resistance R as function of back gate voltage $V_{BG}$ measured on an hBN-SLG-hBN Hall bar structure [see left inset of (a)] before (black trace) and after (red trace) the first rapid annealing step. The right inset shows the corresponding temperature profile. The scale bar of the inset in (a) is 2~$\mu$m. (b) Corresponding conductivity $\sigma$ and (c) carrier mobility $\mu$ as function of the carrier density $n$. (d) $\mu$ extracted within different $n$ intervals [see dashed lines and symbols in panel (b)] after each annealing step as function of the maximum annealing temperature $T_{max}$. (e) Typical Raman spectrum of the structure after the first annealing step. The 2D-peak shows a narrow width of $\Gamma_{2D}\approx 18.5$~cm$^{-1}$.
        (f) Statistical distribution of $\Gamma_{2D}$ over the entire Hall bar structure shown in the inset of panel (a) before (black) and after (red) the first annealing step.
}
\label{fig:MPRs}
\end{figure}

\textit{Fabrication} In Fig. 2 we show an illustration of the technique to fabricate SLG/BLG-hBN heterostructures. In particular, we show the transfer process of graphene on hBN which is based on the method developed by Wang et al. \cite{Wang2013Nov}. In a first step, graphene and hBN with a thickness of around 10-30 nm is exfoliated on two different, highly p-doped Si substrates with a top layer of thermal SiO$_2$. A glass slide with an adhesive tape and a spin coated 1 m thick copolymer (Elvacite 2550) is then prepared similar to Ref. \cite{Zomer2011Dec}. This glass stack is used as a substrate for another hBN flake. In a second step, the glass slide is placed in a mask aligner, where we align the hBN to the graphene on SiO$_2$ and bring both flakes into contact. Due to the strong van der Waals forces between both atomically flat materials (see lower inset of Fig. 2), the graphene attaches to the hBN and can be lifted from the SiO$_2$. Subsequently, in an additional mask aligner step, we place the lifted stack on the hBN resting on SiO$_2$ and release the stack by heating the sample together with the glass slide to 90$^\circ$C. Finally, the sample is cleaned by organic solvents leaving the hBN-SLG/BLG-hBN structure on the SiO$_2$.The sample is then structured by patterning a chromium hard mask using electron beam lithography and a reactive ion etching step based on SF$_6$ plasma. Please note that during this process the sample is heated up to 180$^\circ$C which is why in the following we focus on annealing processes above this temperature. After removing the hard mask in a wet etching process, the contacts are patterned by another electron beam step and evaporation of Au/Cr (60 nm/5 nm) followed by lift off, resulting in one-dimensional contacts at the edges of the sandwich structures as shown by Wang et al. \cite{Wang2013Nov}. An example of a contacted device is shown in the inset of Fig. 3a.

\textit{Encapsulated SLG} In the following, we investigate the properties of an hBN-SLG-hBN heterostructure by transport and Raman spectroscopy measurements before and after thermal annealing. The left inset of Fig. 3a shows an optical image of the investigated Hall bar structure with width $W = 1.9~ \mu$m and distance $L = 2.0~\mu$m between the voltage probes. Furthermore, the temperature profile of the first rapid annealing step is shown in the right inset of Fig. 3a where a peak temperature of $T_{max}$ = 380$^\circ$C is reached after a time of $t$ = 14~min. The annealing is carried out in a home made oven in vacuum at a pressure of $p = 0.2$~mbar. Figure 3a shows the four-terminal resistance $R$ of the device before (black trace) and after (red trace) the annealing in dependence of the back gate voltage $V_{BG}$. After the annealing, $R$ exhibits an increase of its maximum resistance $R_{max}$ from 4.5 to 9.0 k$\Omega$ and a shift in $V_{BG}$ from −2.64 to −2~V, which corresponds to a small change in doping of $\Delta n=4.3\times 10^{10}$cm$^{-2}$. The increase in $R_{max}$ is in agreement with the extracted residual carrier density at the Dirac point, which we extract following Ref. \cite{Couto2014Oct}, obtaining values of $n^*\approx5.0\times 10^{10}$cm$^{-2}$ before and $1.1\times 10^{10}$cm$^{-2}$ after annealing. Furthermore, by comparing the two-terminal with the four-terminal resistance in the electron (hole) regime, we extract a contact resistance $R_c$ of 3.0 k$\Omega$ (2.1 k$\Omega$) before and 2.6 k$\Omega$ (2.8 k$\Omega$) after the annealing step. Here, the different tendencies in the electron and hole regime together with the small change in magnitude leads to the conclusion that the annealing process does not result in a degradation of the contacts. To further extract the carrier mobility $\mu$, we extract the corresponding conductivity $\sigma$ of the device and plot it against the carrier density $n$ (see Fig. 3b) which is related to $V_{BG}$ by $n = \alpha V_{BG}$ where $\alpha = 6.7$~cm$^{-2}\times 10^{10}$ V$^{-1}$ is the lever arm of the back gate extracted from quantum Hall measurements. In the following, $\mu$ is extracted by two different methods, i.e. by calculating the derivative $\mu = ($d$\sigma/$d$n)/e$ and by performing a linear regression. The extracted $\mu$ values are shown in Fig. 3c and d respectively. Figure 3c shows an almost unchanged mobility of 40,000-60,000 cm$^2$/(Vs) before (black) and after (red) the first annealing step for high carrier densities $n > 0.7\times 10^{12}$ cm$^{-2}$. This behavior can be confirmed by the data shown in Fig. 3d where we plot $\mu$ extracted by linear regression against the $T_{max}$ of the corresponding annealing step. Please note that the first annealing step was carried out with the second lowest $T_{max}$ = 380$^\circ$C. For the second annealing we chose $T_{max}$ = 355$^\circ$C. In the subsequent annealing steps, $T_{max}$ was increased for each step. The values extracted in this way confirm the values shown in Fig. 3c. The mobility extracted by linear regression within the interval of $0.7 \times 10^{12}$~ cm$^{-2} < |n| < 2.0 \times 10^{12}$ cm$^{-2}$ (see outer dashed lines in Fig. 3b and symbols) stays almost constant with varying $T_{max}$. Moreover, if $\mu$ is extracted in the range of $0.1 \times 10^{12}$~ cm$^{-2} < |n| < 0.5 \times 10^{12}$ cm$^{-2}$ (see inner dashed lines in Fig. 3b and symbols), the values fluctuate between 60,000 and 100,000 cm$^2$/(Vs) without following a particular trend.

To gain access to more information on sample properties, we perform micro Raman spectroscopy at a laser wavelength of 532 nm. A typical spectrum is shown in Fig. 3e which features an hBN related peak at $\omega_{hBN} = 1366$~cm$^{-1}$, a G-peak at a position of $\omega_{G} = 1582$~cm$^{-1}$ and a 2D-peak at $\omega_{2D} = 2680$~cm$^{-1}$. A striking detail of such a typical Raman spectrum of an hBN-SLG-hBN heterostructure is the narrow 2D-peak width, which is given by $\Gamma_{2D}=18.5$~cm$^{-1}$ in the illustrated spectrum. Compared to the typical full width at half maximum (FWHM) of $\Gamma_{2D} = 30$~cm$^{-1}$ for graphene on SiO$_2$ \cite{Graf2007Feb}, this value is considerably lower and has recently been attributed to the low amount of strain fluctuations within the laser spot size \cite{Neumann2015Sep}. In consequence, monitoring $\Gamma_{2D}$ over the entire sample provides us with additional information about strain fluctuations in the device. The result of this analysis is shown in Fig. 3f where we plot the distribution of $\Gamma_{2D}$. The distribution changes towards lower $\Gamma_{2D}$  values by 1-2~cm$^{-1}$ after the first annealing, indicating a decrease of the overall short range strain fluctuations. Please note that the distribution of $\Gamma_{2D}$  stays unchanged after further annealing steps. Also note that annealing does not lead to any changes in the position of the G and 2D lines due to induced doping or strain.

In summary, the findings discussed above show that the investigated structures are robust against thermal annealing treatments. We observe only slight modifications of the mobility $\mu$, overall doping $n$, contact resistance $R_c$ and absolute strain. However, the residual carrier density $n^*$ and the amount of strain fluctuations are lowered by the first annealing step and remain nearly unchanged in subsequent thermal treatments with temperatures ranging up to T = 550$^\circ$C. These results are in contrast to what has been observed in similar experiments on SLG on SiO$_2$ where $\mu$ strongly degrades by over 50~$\%$ \cite{Cheng2011Feb}. Furthermore, the graphene exhibits a heavy hole doping of over $\Delta n>3.5\times 10^{12}$~cm$^{-2}$ after vacuum annealing \cite{Cheng2011Feb,Ni2010May} which is two orders of magnitudes larger than the observed changes in our study. Finally, the strong broadening of the 2D-line by over $\Gamma_{2D} > 10$~cm$^{-1}$ \cite{Xueshen2013Apr} is in contrast to the observed slight narrowing of $\Delta\Gamma_{2D}\approx 1-2$~cm$^{-1}$ for encapsulated graphene. All these findings show that hBN protects the graphene from oxygen doping, binding to dangling bonds and remote doping by charge traps which are believed to be the dominant causes for doping and the degradation of the mobility on SiO$_2$ \cite{Ni2010May,Hwang2007May,Nistor2012Jul}. Moreover, its atomically flat surface prevents the introduction of strain fluctuations, which can be concluded from the slight narrowing of the Raman 2D-line, and is a major drawback in SiO$_2$ supported devices \cite{Xueshen2013Apr}.

\begin{figure}	
\includegraphics[width=\columnwidth]{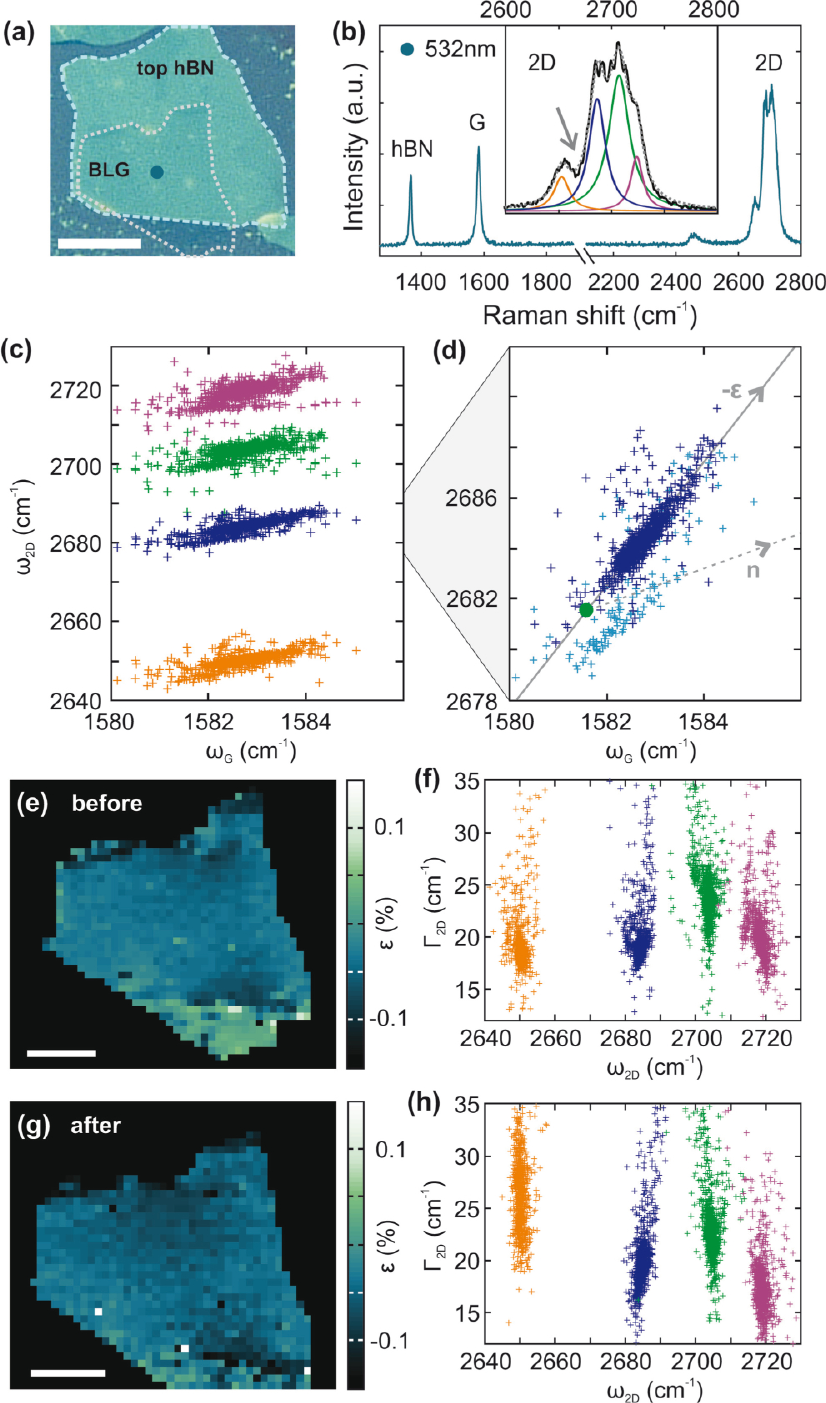}
\caption{
         (a) Optical image of an hBN-BLG-hBN heterostructure. The scale bar represents 10~$\mu$m. (b) Typical Raman spectrum recorded at the position indicated by the blue dot in (a). The upper inset shows a closeup of the 2D-peak together with the extracted four sub-peaks. (c) Correlation of the 2D-sub-peak positions $\omega_{2D}$ [compare colors in panel (b)] to the G peak position $\omega_{G}$ for a Raman map recorded over the entire sample shown in panel (a). (d) Closeup of the $\omega_{G}/\omega_{2D}$ correlation for the second 2D-sub-peak. The data points marked in light blue are taken at positions where the BLG is not covered by the top hBN. Solid and dashed lines show the axis related to strain (-$\epsilon$) and doping ($n$) while the green dot indicates the point of zero strain and zero doping which is used to perform the vector decomposition method \cite{Lee2012Aug}. Panels (e) and (g) illustrate the resulting strain values before and after the thermal annealing process in Ar/H$_2$. (f) and (h) show the corresponding extracted 2D-sub-peak widths $\Gamma_{2D}$  for all four sub-peaks versus the respective sub-peak position $\omega_{2D}$. The scale bars in panel (e) and (g) represent 5~$\mu$m.
}
\label{fig:MPRs}
\end{figure}

\textit{Encapsulated BLG} In the following, we extend the annealing study to an hBN-BLG-hBN heterostructure. An optical image of such a structure is shown in Fig. 4a. The dashed lines highlight the outlines of the top hBN (light blue) and BLG (gray) flake, while the entire structure is placed on a large bottom hBN flake. A lower part of the BLG flake is not covered by the top hBN layer, as can be seen in Fig. 4a. Figure 4b shows a typical Raman spectrum recorded at the position indicated by the blue dot in Fig. 4a before the annealing process. The spectrum shows the characteristic fourfold splitting of the 2D-peak into four sub-peaks as illustrated by the inset in Fig. 4b. Similar to the Raman spectra of SLG the 2D-sub-peaks exhibit an exceptionally small FWHM, which-manifests-itself in the dip of the Raman signal between the two leftmost sub-peaks (see arrow in the inset of Fig. 4b), and has not been observed in Raman spectra with a laser wavelength of 532 nm for BLG before \cite{Graf2007Feb,Ferrari2013Apr,Ferrari2006Oct,Zabel2012Feb,Malard2007Nov,Yan2008Sep}. As a consequence, by fitting four Lorentzians to the 2D-peak, we can identify the positions of the 2D-sub-peaks with a high precision. The results of such an analysis are plotted in Fig. 4c where we correlate the four $\omega_{2D}$ to the G-peak position $\omega_{G}$ of a Raman map recorded over the entire sample shown in Fig. 4a before annealing. The colors of the data points correspond to the four different sub-peaks of the same color highlighted in the inset of Fig. 4b. Four distinct data clouds are visible. To obtain more information on the properties of the BLG, we focus on one of the data clouds (i.e. the second sub-peak data points) as highlighted in Fig. 4d. Here, the light blue data points correspond to spectra measured at positions where the BLG is not covered by the top hBN layer. Having obtained precise values for $\omega_{2D}$ for each sub-peak, we are able to perform the vector decomposition method developed by Lee et al. \cite{Lee2012Aug}, which was previously only applied to SLG. By correlating $\omega_{2D}$ and $\omega_{G}$, it is possible to extract local strain and doping values of the graphene flake. To perform the analysis, we choose the point of zero doping and zero strain to be located at $\omega_{G,0} = 1581.6$~cm$^{-1}$ and $\omega_{2D,0} = 2681.5$~cm$^{-1}$ (see green dot in Fig. 4d) \cite{Yan2008Sep}. Furthermore, for the strain dependent shift, we use the experimentally determined values of $\Delta\omega_{2D}/\Delta\omega_{G}=2.45$ (see solid  line in Fig. 4d) which corresponds to the Gruneisenparameter for the G-peak of $\gamma(G)=1/\omega_G  \cdot\partial\omega_G/\partial\epsilon=1.8$ as extracted by Zabel et al. \cite{Zabel2012Feb} in the presence of biaxial strain. In case of doping, the extraction of doping values for BLG is not sensible, since the non-monotonic behavior of the optical phonon anomaly \cite{Yan2008Sep} prevents an assignment of the doping induced G-peak shift. Hence, we are not going to discuss the doping related G-peak shift in the following. Please note that the dashed line in Fig. 4d illustrates the doping axis for SLG \cite{Lee2012Aug} and should only serve as a guide to the eye.

The results of the projection on the strain axis before and after thermal annealing at a maximum temperature of $T_{max}$ = 375$^\circ$C are shown in Figs. 4e and g, respectively. In both maps, the extracted large scale strain values vary by less than 0.2~$\%$ which is in good agreement with recent findings on encapsulated SLG in hBN \cite{Neumann2015Sep}. Moreover, the overall strain values do not change significantly after the annealing step. Another interesting detail is the apparent difference in strain values for BLG which is not covered by the hBN with respect to the encapsulated BLG. Figure 4e shows a strong increase of the strain values in the uncovered area. This finding is also reflected in the second light blue data cloud in Fig. 4d which corresponds to spectra recorded on the uncovered area and is clearly shifted towards lower $\omega_{2D}$  values. In Figs. 4f and h we show the correlation of the 2D-sub-peak width with the 2D-sub-peak position. While the first 2D-sub-peak (orange data points) shows a tendency of an increased $\Gamma_{2D}$, we observe a decrease in $\Gamma_{2D}$ for all the three remaining sub-peaks, in agreement with the experiment on SLG described above. This finding leads us to the conclusion that BLG in general shows the same tendency as SLG, where we find a decrease in $\Gamma_{2D}$ pointing to a lower amount of short range strain fluctuations \cite{Neumann2015Sep}.

Please note that the characterized BLG structure was subsequently patterned and contacted and transport measurements are shown in Ref. \cite{Engels2014Sep}. The extracted carrier mobility of around $\mu$ = 40,000-50,000 cm$^2$/(Vs) and low residual doping values of $n^* = 4.3 \times 10^{10}$~cm$^{-2}$ are again underlining the high quality of these state of the art BLG-hBN heterostructures.

\textit{Conclusion} In conclusion, we presented a thermal annealing study on hBN-encapsulated SLG and BLG heterostructures. We found that these structures are robust against thermal cycling. While an initial annealing step might lower residual doping and strain fluctuations of the sample, properties such as the charge carrier mobility, overall doping, and contact resistance stay nearly constant up to temperatures of 525$^\circ$C. Furthermore, the low strain fluctuations and the resulting small line width of the Raman 2D-sub-peaks of BLG heterostructures allowed us to precisely extract the position $\omega_{2D}$ of these peaks. In turn, this allowed us to perform the vector decomposition method on BLG and extract strain values which underline the robustness of encapsulated BLG against thermal annealing. All findings are in contrast to graphene on SiO$_2$ substrates, where thermal annealing leads to a significant degradation of the discussed material properties. In consequence, our results show that hBN-graphene heterostructures are particularly interesting for applications in the presence of extreme thermal conditions.

\textit{Acknowledgements} We thank R. V. Gorbachev for helpful discussions on the fabrication process. Support by the HNF, JARA Seed Fund, the DFG (SPP-1459 and FOR-912), the ERC (GA-Nr. 280140), and the EU projects Graphene Flagship (contract no. NECT-ICT-604391) are gratefully acknowledged.

\end{document}